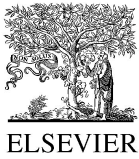

# A Decision Matrix and Monitoring based Framework for Infrastructure Performance Enhancement in A Cloud based Environment

Mansaf Alam [a], [*] Kashish Ara Shakil [a]

[a] *Department of Computer Science, Jamia Millia Islamia( A Central University), New Delhi-25, India*

**Abstract**

Cloud environment is very different from traditional computing environment and therefore tracking the performance of cloud leverages additional requirements. The movement of data in cloud is very fast. Hence, it requires that resources and infrastructure available at disposal must be equally competent. Infrastructure level performance in cloud involves the performance of servers, network and storage which act as the heart and soul for driving the entire cloud business. Thus a constant improvement and enhancement of infrastructure level performance is an important task that needs to be taken into account. This paper proposes a framework for infrastructure performance enhancement in a cloud based environment. The framework is broadly divided into four steps: a) Infrastructure level monitoring of usage pattern and behaviour of the cloud end users, b) Reporting of the monitoring activities to the cloud service provider c) Cloud service provider assigns priority according to our decision matrix based max-min algorithm (DMMM) d) Providing services to cloud users leading to infrastructure performance enhancement. Our framework is based on decision matrix and monitoring in cloud using our proposed decision matrix based max-min algorithm, which draws its inspiration from the original min-min algorithm. This algorithm makes use of decision matrix to make decisions regarding distribution of resources among the cloud users.

## 1. Introduction

Cloud computing is an emerging technology which has a potential for causing a paradigm shift in the way computing resources are used with the aim of leading to a major reduction in costs incurred in developing an application, and thus provides several advantages at both the end user as well as the service provider level. Cloud computing enables, computing resources to be made available as a service over the network and thus provides its users with an added advantage of geographical independence along with an elevated level of agility, scalability and dynamic access. It also provides computing resources in a virtualized manner.

*1.1 Forms of Cloud Computing*

There are different forms in which cloud is available:

*a) Public Cloud:* Public cloud is one in which computing resources such as applications, servers, storage, computing infrastructure and other services are made available to the general public. These services are either available freely to the users or provided on a rental basis. Amazons EC2 and Google AppEngine are some of the public cloud services available at present.

*b) Private cloud:* Private cloud is a cloud set up which is managed for a particular organization. Since a private cloud is set up only for a particular organization therefore the users of private cloud are also limited to the organization level. Therefore, users of a private cloud will be employees within the organization who are authorized to access the cloud services. For Example organizations such as HP and Microsoft have a private cloud of their own.

*c) Community cloud*: Community cloud is a cloud set up where the cloud is shared amongst multiple organizations having common interests such as common security, resources and policy concerns. Sharing of a cloud amongst multiple organizations provides varied amount of cost benefits to the organizations. For Example Google Gov is a community cloud.

* Corresponding author. Tel: Tel: 91-9810650497
E-mail: malam2@jmi.ac.in

*d) Hybrid cloud:* Hybrid cloud is a cloud which is a composition of two or more different types of clouds. It can be a combination of public and private cloud or community and public. Thus it allows organizations to manage their resources either internally i.e. within the organization or externally i.e. outside the organization. For example Amazons EC2 can be used as a hybrid cloud.

*1.2 Characteristics of cloud*

Few key characteristics of cloud based computing according to NIST are [5]:
a) *On demand self-service*: Cloud computing enable users to use computing resources such as servers and storage depending on their requirement. Thereby using resources on demand without any intermediate party.
b) *Broad network access*: Provides computing resources as services over a network and these resources can be accessed through a standard interface.
c) *Resource pooling*: Cloud computing involves pooling of several resources. The resources are pooled and made available to users as a service. It allows same resources to be allocated and reallocated amongst multiple users.
d) *Rapid elasticity*: It allows resources to be scaled up and down in an elastic manner to the users. Thereby providing users with an illusion of unlimited resource availability.
e) *Measured service*: Cloud computing allows computing resources to be made available to users' as a service in a pay per use manner i.e. the users pay only for the part of service they have used.

Cloud computing can also act as an efficient way to manage data, [15] describes a technique for data management in cloud using k-median clustering technique. Though cloud computing promises to be a highly profitable technology with several advantages but it is not free from limitations such as security etc., Thus in order to ensure if the cloud is performing as per the promise of service provider it is necessary to monitor the infrastructural resources in cloud [6]. Performance monitoring in cloud faces many challenges and the current techniques available for diagnosis cannot solve these challenges [12]. In this paper we first present the need for monitoring data in cloud, categories of performance monitoring in cloud along with its benefits and then propose our framework which is based on decision matrix and monitoring for enhancing the infrastructure performance of cloud. We have also proposed an algorithm called DMMM algorithm which draws its motivation from the original min-min algorithm.

## 2. Related Work

In cloud computing resources are made available to a user as a service over the network. The cloud end users pay only for the part of service they are interested in i.e. they make use of the concept of pay per use. Since the client is paying for the services he/she is using therefore quality of service received is a prerequisite for the clients. Similarly to stay in business service providers also have to insure that the service delivered by them has an edge over other competitors in terms of quality, performance etc. Thus, monitoring performance and thereby improving the quality of service is highly desirable.

Runtime model for cloud monitoring (RMCM) organizes data gathered from different monitoring techniques [1]. It provides a feasible and effective approach for providing a balance between overhead occurring during runtime and management of monitoring of resources. In [2] an evaluation of comparison of HPC platforms with Amazon EC2 is performed and it indicates that EC2 is six times slower than a midsized Linux cluster and thus it has a poor impact on performance of cloud computing platform. A performance model based on a monitoring framework can be used for guaranteeing the performance of cloud applications [3].This model collects runtime monitoring data on a real cloud such as PaaS. Complex event processing can be used for monitoring application level performance [9].In [9] an event based approach is used for applications based on cloud and according to it each system in cloud generates some event. Therefore each system on a cloud network is called as event emitter. The events generated are processed through event correlation technique using CEP. The Cloud scale framework can then be used for acquiring and releasing resources dynamically in a cloud environment.

Networking and cloud computing are now related terms as a network is responsible for provisioning of resources in cloud and the quality of network has an impact on performance of cloud services[11]. So, cloud computing and networking can both be coupled together to get users perspective of performance of cloud services. SOA can be used for composition of cloud and network together. A service oriented architecture based on network virtualization is described in [11] which clubs networks and cloud provisioning system together for improving performance of cloud. Cloud trust [13] is a trust based model for evaluating the performance of cloud environment on the basis of its several attributes. It uses rough set and IOWA operator as adaptive modelling tools for application to trust based data mining and thereby leading to knowledge discovery. Windows base state monitoring framework [14] can be used as an efficient monitoring framework. It provides a 50-90 percent reduction in communication overhead as compared to the other instantaneous ones and is also more scalable.

## 3. Need for Monitoring In Cloud

Monitoring cloud performance involves monitoring from two point of views i.e. there are two beneficiaries of cloud computing: Cloud service provider and cloud end users [4].Cloud service providers make sure that the services are well allocated and also monitor the resources in order to meet up to the demand of the end users as well as monitor the network for any illegal or malicious attacks. End users also have to ensure the quality of service they are receiving i.e. whether they are getting value for their money or not. They also need to be made sure about the security of their data once it is put in the cloud infrastructure. They need to monitor whether their data is not being misused by any malicious attackers etc. The key activities for which resource monitoring is important described by [7] are:
  a) *Capacity and Resource planning*: Cloud capacity and resource planning is essential for service providers in order to meet up to the quality of service guaranteed by them in SLA and to insure cloud services are provided in a smooth fashion.
  b) *Capacity and Resource Management*: after the resources have been carefully planned next important task is resource management in order to insure availability of resources at all times.
  c) *Data Center Management*: Since the cloud data is hosted in a data center therefore it is necessary to manage data centers .This involves tasks such as monitoring and data analysis.
  d) *SLA management*: When auditing of cloud services is done it is monitored against its compliance to the corresponding SLA.
  e) *Billing*: Since cloud computing provides resources in a pay per use manner therefore it is instrumental for both the service providers as well as the end users to monitor the billing of their respective activities.
  f) *Troubleshooting*: It is necessary for both the service providers as well as the end users to understand the cause of their problems whether it's a problem at their end or at network level.
  g) *Performance Management*: It is necessary for cloud users to monitor their cloud tasks regularly especially if it involves real time applications where time is a constraint.
  h) *Security Management*: Security in cloud is an issue which must be taken into consideration when moving to cloud. Therefore it is necessary to monitor the security of sensitive information in cloud.

## 4. Categories of Performance Monitoring In Cloud

Performance monitoring of cloud can be classified into 2 categories [8]:

### 4.1. Infrastructure Performance

Infrastructure performance of cloud involves performance of various resources that are provided as a service in cloud. It includes performance of servers, network, storage etc. Infrastructure response time can be used for finding out the performance of cloud infrastructure. It is defined as the time between submission of a request and the time for cloud infrastructure to complete the request. Therefore for measuring performance of cloud infrastructure multiple platforms must be supported along with calculation of infrastructure response time and identification of resources being used.

### 4.2. Application Performance

Application performance management involves monitoring of resources in cloud that provide support to application program performance. It is of interest to the cloud consumers. Application Response time is used for measuring the performance of application. Application performance monitoring also involves monitoring of database, end user experience monitoring, monitoring of application and web servers and SLA management [10].

## 5. Benefits of Performance Monitoring In Cloud

There are various benefits of monitoring performance in cloud such as:
  a) *Identification of shortage of resources*: This is particularly beneficial for organizations where organizations can keep a track of performance of their applications and also monitor if the resources which are currently available with them are able to handle peak volumes of resource requirements and if the resources are not optimal they can increase or decrease their number accordingly.
  b) *End Users keep an eye on the value for their money:* with performance monitoring at client level cloud users can benefit by making sure they are getting their resources or services as per their demand.
  c) *Forecast and prevent performance issues* [10]: Since performance monitoring enables cloud organizations to watch the resource utilization in cloud .This enables them to predict weather crisis in a timely manner.
  d) *Dynamic support for changes in organizational demands* [10]: Since performance and business monitoring is done by organizations from end to end. This enables organizations to provide a dynamic support against business decisions such as addition or removal of resources.
  e) *Optimal Spending decision support*: Since cloud computing is a metered technique monitoring its performance provides organizations with the facility which aids them in making decisions regarding where to spend and how much.

## 6. Proposed Framework

According to a survey conducted by "451 market research" there will be a 37 percent growth rate (compound annual growth rate) in IaaS by 2016. Thus Infrastructure level performance is an issue which must be taken into account and therefore its performance enhancement is a decisive task. Our proposed framework aims at improving the infrastructure performance of cloud using a decision matrix based max-min algorithm (DMMM). It uses a combination of decision matrix based max-min algorithm and monitoring techniques to improve infrastructure performance of cloud. Our approach proves to be a lucrative one for cloud performance enhancement, as it leads to a win-win situation for both the service providers as well as the cloud users. It's a win-win situation because it leads to mutual benefits for both the service providers as well as cloud users leaving them both satisfied and content. Service providers benefit by distributing resources as per their business strategies and availability, on the other hand clients also benefit by getting improved services. Fig 1 given below provides a brief outline of our proposed framework.

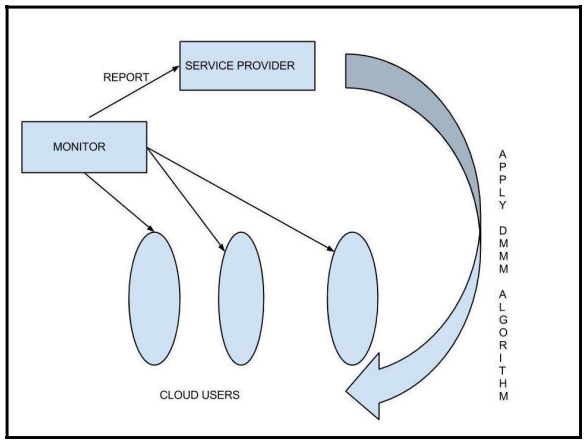

Fig 1. Priority based infrastructure performance enhancement/Proposed Framework

Cloud computing allows computing resources to be accessed as a service irrespective of the geographical locations .It allows computing resources to be accessed in a dynamic, elastic and a highly scalable manner. The cloud users tend to use computing resources as per their requirements in a dynamic manner. Sometimes they use lesser resources and sometimes more depending upon their usage and workload demand. Our Framework takes advantage of this fluctuation in the usage of resources by monitoring the usage pattern of customers. Our Framework for performance enhancement is broadly divided into the following steps or stages:
  a) Infrastructure level monitoring of usage pattern and behaviour of the cloud end users.
  b) Reporting of the monitoring activities to the cloud service provider
  c) Cloud service provider assigns priority according to our decision matrix based max-min algorithm (DMMM) which will be discussed later.
  d) Providing services to cloud users leading to infrastructure performance enhancement.

*6.1 Infrastructure level monitoring of usage pattern and behaviour of the cloud end users*

Cloud users use computing resources in a non uniform manner depending upon their requirements. Therefore, at this stage an infrastructure level monitoring of the usage of resources and servers is done to find out the usage pattern of the cloud end users i.e. a monitoring of the activities of cloud users relating to their usage pattern is done. The usage pattern easily deciphers the hours of peak demand of resources, the kind of resources being used, the duration during which least resources are used or the time duration during which the users are least active or in dormant stage. For example the graph given below (Fig 2) illustrates the usage pattern of three resources Resource 1, Resource 2 and Resource 3 by four customers of an organization.

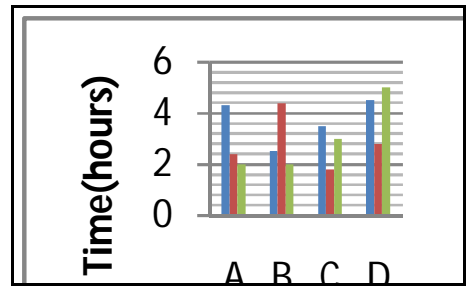

Fig 2: Usage Pattern of Customer's of an Organization

This usage pattern can prove to be highly beneficial to the infrastructure service providers as, it gives the service providers a clear picture of the amount of resources being used and the duration of time for which the resources are being used by a particular client.

*6.2. Reporting of the monitoring activities to the cloud service provider*

After the usage pattern has been monitored and data regarding the usage pattern has been generated, (which can either be in graphical form or tabular), this information is then reported to the cloud service providers. After the receipt of this information regarding the usage pattern, the experts at the cloud service providers than asses this information in order to carry out their tasks of planning the allocation and reallocation of resources and thereby distribute and redistribute their resources accordingly. They also get information about the peak usage hours and also the hours when the demand for a resource is minimal. This can prove to be highly beneficial for organizations, as they can expand their business by attracting more probable clients during less active hours.

*6.3. Cloud service provider assigns priority according to our decision matrix based max-min algorithm (DMMM)*

Cloud computing involves variety of systems, resources, tasks and data. Since cloud computing involves a variety of systems therefore scheduling of tasks is now considered as an important issue. This stage in our framework is meant for assigning resources to customers.

After the information about the usage pattern is reported to the cloud service providers, the cloud users then assigns priority to its end users using our proposed max min algorithm.

DMMM draws its inspiration from the original Min-Min scheduling algorithm. According to min-min algorithm if there are 'n' numbers of tasks i.e. T= {t1, t2…tn} and 'm' number of resources i.e. R= {R1, R2…….Rm}, where R represents set of resources and T set of tasks. Then min {T} is selected and assigned to min{R} until all the tasks have been assigned resources. The Min-Min algorithm is now also being used in distributed environments and infrastructures like cloud.

*6.3.1 DMMM(Decision matrix based max-min algorithm)*

Let $T=\{t_1, t_2 \ldots t_n\}$ be a set of tasks of cloud users who have to be assigned resources $R=\{r_1, r_2 \ldots r_m\}$ and let $X_{ij}$ be a value calculated from decision matrix having $v_1, v_2 \ldots v_k$ as the outputs such that $X_{ij}$ = maximum value of $v_1, v_2 \ldots v_k$. Our algorithm selects the resource with value of $X_{ij}=\max(v_1, v_2 \ldots v_k)$ and assigns this resource to task having minimum execution time.
Our proposed DMMM algorithm is given by Fig 3

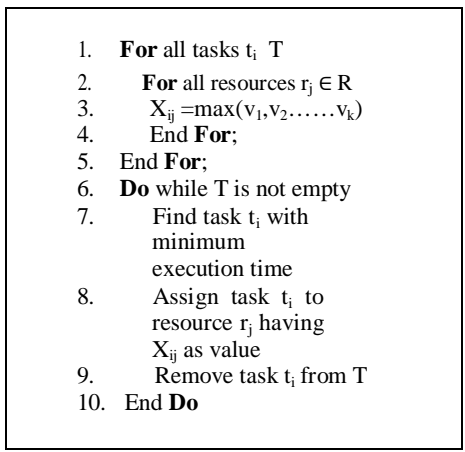

1. **For** all tasks $t_i$ T
2.    **For** all resources $r_j \in R$
3.    $X_{ij} = \max(v_1, v_2 \ldots v_k)$
4.    End **For**;
5. End **For**;
6. **Do** while T is not empty
7.    Find task $t_i$ with minimum execution time
8.    Assign task $t_i$ to resource $r_j$ having $X_{ij}$ as value
9.    Remove task $t_i$ from T
10. End **Do**

Fig 3. DMMM algorithm

Our proposed algorithm first finds out the resources having maximum value amongst all the tasks, it also finds out the task that requires minimum time for execution. It then assigns the resource having maximum value to the task having minimum execution time.

A decision matrix can be used as an effective tool for making various decisions and also for comparing different available alternatives. A cloud infrastructure provider can take advantage of decision matrix for distributing resources amongst the various cloud users. One way of achieving this is by assigning a priority value to each of the customers depending on the type of client such as a day to day client, important client, privileged client etc. Therefore in order to meet this end we have used a decision matrix to calculate the value of Xij, our decision matrix is based on priorities and criteria's decided by the service providers.

*6.4. Providing services to cloud users leading to infrastructure performance enhancement*

After the priorities are assigned by the service providers to the cloud users according to our proposed DMM algorithm the cloud service providers then distribute their resources to the end users as per the assigned priority i.e. the key customer or the privileged customer with the highest priority is provided services first as soon as they demand followed by other users depending on their respective priorities. Providing key customers services with high priority leads to creating of a win-win situation for both the cloud service providers as well as the cloud end user's or customers.

## 7. ILLUSTRATIONS AND IMPLEMENTATIONS

In order to demonstrate our algorithm we will use an illustrative example, We assume that we have four cloud users with tasks t1,t2 ,t3,t4 and three resources r1,r2 and r3 .Now these resources are to be distributed amongst the given tasks. The duration of each task is given by Table 1.

Table 1-Execution Time of each task

| Tasks | Execution time(time units) |
|---|---|
| t1 | 15 |
| t2 | 20 |
| t3 | 10 |
| t4 | 5 |

As per the criteria's decided by the cloud service providers based on their preferences, priorities and other information's deduced from the usage pattern, a decision matrix will be drawn that will enable them in distributing resources amongst the different cloud users. For example for allocation of resource r1 the criteria would based on type of users i.e. we will associate a priority value for each of the users. For illustrative purpose we have used '4' to represent highest priority value and '1' as a representative of lowest priority value. Suppose we have four types of users: Benefited users, important users, Casual users and lesser privileged users. We have associated a priority value with each of the users. The different types of users and their associated priorities are given by Table 2.

As per the criteria's decided by the organizations for resource distribution, a decision matrix is drawn also given by Fig 4.

| USER TYPES | | | | | |
|---|---|---|---|---|---|
| | | BENEFITED | IMPORTANT | CASUAL | LESSER PRIVILEGED |
| CRITERIA'S | WEIGHT | PRIORITY=4 | PRIORITY=3 | PRIORITY=2 | PRIORITY=1 |
| CRITERIA C1 | 1 | 1*4 | 1*3 | 1*2 | 1*1 |
| CRITERIA C2 | 2 | 2*4 | 2*3 | 2*2 | 2*1 |
| CRITERIA C3 | 3 | 3*4 | 3*3 | 3*2 | 3*1 |
| TOTAL | Vi | 24 | 18 | 12 | 6 |

Fig 4. Organizations Criteria based Decision Matrix

Table 2- User Types and Their Priorities

| USER TYPES | PRIORITY |
|---|---|
| Benefited users | 4 |
| important users | 3 |
| Casual users | 2 |
| Lesser privileged users | 1 |

Now on the basis of this decision matrix we will calculate Xij i.e. the maximum value of Vi Xij=max {24, 18, 12, 6} =24

Hence as per our algorithm task t4 is assigned to resource r1 .Similarly, we will use a decision matrix to assign resources to the remaining tasks.

## 8. Advantages of the Proposed Framework

Our proposed framework aims to provide several benefits to both the cloud users as well as the service providers. Following are some of the benefits of our approach:
1) *Infrastructure Performance enhancement*: Since resources are assigned by users cloud service provider after careful monitoring of the cloud users .Therefore it leads to infrastructure performance improvement.
2) *Win-Win condition for both:* Both cloud users and service providers benefit. Service providers benefit by getting better control over their resource distribution. Client benefit by getting better services.
3) *Better Services:* Since cloud users can now provide their services in a more manageable manner. Hence quality of service also improves.
4) *More value for money:* Since services are provided by service providers on the basis of a decision matrix. Therefore clients can choose what level of quality do they expect and thereby get better value for their money.
5) *More Manageable business activities:* It mabusiness more manageable as the service providers can make decisions as per their own convenience.
6) *Better business opportunities for organizations:* Since now the distribution of resources is within the reach of service providers therefore, this framework would open gates for newer opportunities for service providers

## 9. CONCLUSION AND FUTURE WORKS

Cloud computing is a very fast developing technology which is developing at an explosive rate. The cloud comes in different forms such as private, public, hybrid and community. It has several overwhelming characteristics such as on demand self-service, high scalability and rapid elasticity. Cloud infrastructure is very different from traditional computing infrastructures therefore monitoring of data in cloud is also distinguished. Cloud monitoring is important for activities like capacity and resource management, SLA management, Data center management etc. Infrastructure performance monitoring and Application Performance monitoring are two categories of performance monitoring in cloud. Various benefits of performance monitoring in cloud include identification of shortage of resources, end users keep an eye on the value for their money, forecast and prevent performance issues and dynamic support for changes in organizational demands. In this paper we have proposed a framework for enhancing infrastructure performance in cloud. This framework uses monitoring techniques for monitoring the activities of the users and then calls DMMM which is our proposed algorithm and thereby distributes resources amongst the various cloud users. DMMM draws its motivation from the original min-min scheduling algorithm and uses a decision matrix for assignment of resources to the cloud users. The proposed framework has several benefits such as Infrastructure Performance enhancement, Win-Win condition for both client and service providers, Better Services, more Value for money, more manageable business activities and better Business opportunities for organizations. This paper has developed the framework keeping infrastructure performance improvement in mind but this concept of using a decision matrix can be extended further to the application level and thus ,eventually to the client level as well. Our algorithm seeks its motivation from Min-Min algorithm but this approach can be extended to other similar scheduling algorithms such as max-min, round robin and genetic algorithms.

## Acknowledgment

Kashish Ara Shakil would like to express her gratitude to Dr. Mansaf Alam for his esteemed guidance and being a continuous source of inspiration during the writing of this paper.